\documentclass[twocolumn,showpacs,epsfig,pra]{revtex4}
\usepackage{graphicx}
\usepackage{amssymb}
\usepackage{color}

\begin{document}

\draft
\title{Coupled non-identical microdisks: avoided crossing of energy levels and unidirectional far-field emission}
\author{Jung-Wan Ryu$^{1}$}
\author{Soo-Young Lee$^{2}$}
\author{Sang Wook Kim$^{1}$}
\email{swkim0412@pusan.ac.kr}
\affiliation{$^1$Department of Physics Education, Pusan National University, Busan 609-735, Korea}
\affiliation{$^2$Department of Physics and Astronomy, Seoul National University, Seoul 151-742, Korea}
\date{\today}

\begin{abstract}
We investigate two coupled microdisks with non-identical radii focusing on the parametric evolution of energy levels and the unidirectional far-field emission. We show that the evolution of energy levels is characterized by the avoided crossing intrinsically associated with the exceptional point or the non-Hermitian degeneracy. These spectral properties explain highly asymmetric near-field intensity pattern of the resonance mode. The observed unidirectional far-field emission is shown to be understood by considering the forbidden inter-disk coupling in the ray picture induced by the frustrated total internal reflection near the closest point between two disks when the inter-disk distance is small enough.
\end{abstract}
\maketitle
\narrowtext

\section{Introduction}

Dielectric circular microdisks have been intensively studied because they provide small, high Q factor, and ultralow threshold laser cavities \cite{McC92,Yam93}. All these advantages come from the so-called whispering gallery modes (WGM) based upon total internal reflection of the ray at a dielectric surface. For device applications, coupled microdisks have recently attracted much interest in the context of photonic molecules \cite{Bay98,Muk99,Har03,Nak05,Ish05,Bor06_1}. Coupled two identical microdisks resemble diatomic molecules in that they show the energy level splitting, the formation of bonding (symmetric) and anti-bonding (anti-symmetric) states, and the oscillatory behavior of energy levels on the distance between two disks \cite{Ryu06,Smi03}. All these properties are generic in coupled optical systems such as the coupled bits of dielectric matter \cite{Bur89} and the coupled two dielectric spheres \cite{Cha01,Gom04,Ng05,Kar07}. It is then natural to ask what happens if the two disks are not identical. The research in this direction is rather rare; the avoided level crossing as a function of the ratio between two radii of the disks has been recently performed numerically \cite{Bor07} and experimentally \cite{Nak05,Ben08}.

A dielectric microdisk is an {\em open} system in the sense that the ray with the incident angle smaller than the critical angle refractively escapes from the cavity. Energy levels of open quantum systems have several distinct properties compared with those of closed ones; first of all the system is described as {\em non-Hermitian} Hamiltonian, so that the eigenvalues are complex and the eigenstates form a non-orthogonal set. In particular the behavior of two interacting energy levels as varying external parameters are mainly governed by the so-called exceptional point (EP), where the complex eigenvalues of the corresponding levels coalesce, i.e. forming the degeneracy of complex eigenvalues \cite{Hei99,Hei00,Hei04,Wie06l,Mul08}. The EP has some noticeable characteristics. To observe it at least two independent external parameters are necessary (in more technical terms it is a codimension-two object). The eigenvalues return to their initial values only after making adiabatic parameter change encircling the EP {\em twice}, which exhibits non-trivial topology in a parameter space. The EP was experimentally observed in microwave cavity \cite{Dem01,Dem03}, and has been recently investigated in a dielectric chaotic cavity \cite{Lee08,Wiersig08a}. In the coupled non-identical microdisks (CNM), two control parameters naturally exist; namely the distance between two disks and the ratio between two radii, which allows us to explore the EP. We will show that the EP plays an important role in describing energy level evolution of CNMs, in particular the avoided level crossing.

A disadvantage of a single dielectric microdisk is that the output emission is isotropic due to its circular symmetry. The directed emission is an important ingredient for device applications. An obvious solution is to break the rotational symmetry. Deformed microcavities with various shapes such as ellipse, quadrupole, stadium, spiral, and so on have been extensively studied in this regard \cite{Cha96,Noc97,Gma98,Kim07}. It is the ultimate goal in this direction to achieve the unidirectional output emission. So far several ideas to realize it have been proposed. When only one symmetry axis of reflection exists, the unidirectional emission, if any, takes place along that direction. However, we emphasize that this statement is neither sufficient nor necessary. The unidirectional emission has been reported in the cavity with rounded triangular \cite{Kur04}, annular \cite{Wie06}, and limacon \cite{Wie08} shapes which all exhibit only one symmetry axis. In particular, it was shown that the avoided level crossing plays a crucial role in explaining the mode in an annular cavity with both high Q and the unidirectional emission. In a limacon cavity the unidirectional emission originates from the ray dynamics following unstable manifolds. In the cavity with spiral shape the unidirectional emission was reported in experiment \cite{Che03}, while the nature of the excited mode is still under debate \cite{SYLee04,Altman08}. Note that the spiral cavity has no symmetry axis. Recently it was found that the bidirectional emission can be obtained in the coupled two identical microdisks by choosing appropriate parameters \cite{Ryu06}. Therefore, one expects that breaking the reflection symmetry of the coupled disks, i.e. using {\em nonidentical} disks, opens the possibility to have the unidirectional emission. Quite recently the single-mode lasing with narrow divergent directional emission in a coupled ring laser has been reported in experiment \cite{Sha08}.

\begin{figure}
\begin{center}
\includegraphics[width=0.45\textwidth]{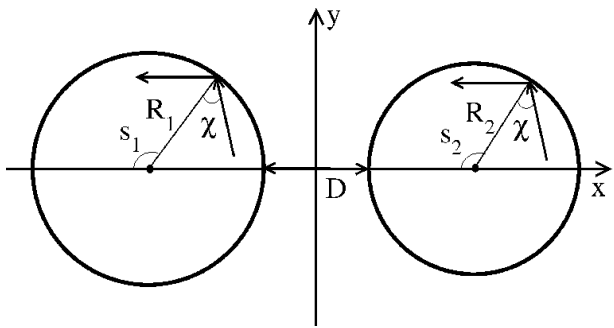}
\caption{The coupled microdisks with radii $R_1$ and $R_2$. $\chi$ and $s$ represent the angle of incidence and the position at the collision of the ray on the boundary, respectively, which is utilized to describe the ray dynamics in Sec~III.}
\label{fig1}
\end{center}
\end{figure}

In this paper, first, we investigate the energy level evolution of CNMs focusing on the avoided level crossing and the characteristics of EPs, especially nontrivial topological structure around the EPs in the parameter space. In fact, we find that the CNM provides an ideal system to investigate the properties of EPs. Varying the ratio between radii of two disks the energy levels show series of typical avoided level crossings. Except the region around the avoided crossing each eigenmode is spatially localized at only one of the two disks. Second, it is shown that the CNM supports a high Q mode simultaneously exhibiting the unidirectional far-field emission. The unidirectional emission is ascribed to short time ray dynamics across the two disks mediated by the {\em frustrated} total internal reflection, where the spatial localization of the eigenmode mentioned above plays a crucial role.

In Sec.~II, we investigate the parameter-dependent complex eigenvalues of the CNM focusing on the avoided resonance crossing, the EP, and the associated near field intensity patterns. In Sec.~III, the ray dynamics of the CNM is studied. In particular the inter-disk coupling mechanism is discussed in detail. In Sec.~IV, we show that the unidirectional emission can take place in the CNM, and explain why it occurs by considering the ray dynamics mediated by the frustrated total internal reflection. Finally, we summarize the paper in Sec.~V.

%%%%%%%%%%%%%%%%%%%%%%%%%%%%%%%%%%%%%%%%

\section{Avoided resonance crossings in coupled non-identical microdisks}

%\subsection{Two coupled microdisks}
Figure~\ref{fig1} shows the coupled microdisks that we consider here. There are two control parameters, namely the interdisk distance $d=D/R_2$ and the ratio between two radii $r=R_1/R_2$, where $R_1$ and $R_2$ represent the radii of the left and the right disks, respectively. The indices of refraction of the dielectric microdisks and the material outside are denoted as $n_{\rm in}$ and $n_{\rm out}$, respectively. We fix $R_2=1$ and set $n=n_{\rm in}/n_{\rm out}=2$. The resonance modes are obtained from solving the Helmholtz equation
\begin{equation}
[\nabla^2 + n^2({\bf r}) k^2]\psi = 0,
\label{helm}
\end{equation}
by using the boundary element method \cite{Wie03}. Here, we consider only TM polarization. Due to the reflection symmetry with respect to the x axis the solutions of the Helmholtz equation are split into two parts depending on their parity, where only those with even parity are considered without loss of generality.

\begin{figure}
\begin{center}
\includegraphics[width=0.44\textwidth]{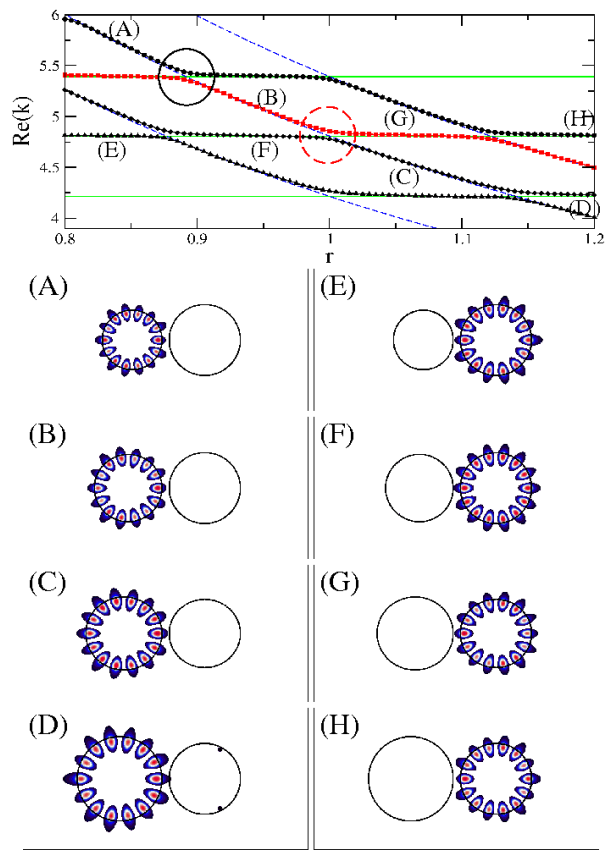}
\caption{(color online). The dots and the thin lines in the upper panel represent the real eigenvalues of the resonances for various $r$ for $d=0.2$ and $d=\infty$ (no coupling), respectively. See the text for the solid and the dashed circles. (A)-(H) show the near field intensity patterns at the specific $r$'s indicated in the upper panel.}
\label{fig2}
\end{center}
\end{figure}

The upper panel of Fig.~\ref{fig2} shows the real part of the wave number $k$ of the resonance modes for various $r$'s with $d$ fixed ($d=0.2$). Two distinct mode groups are clearly identified except near the avoided crossings; the modes horizontally aligned denoted as R-mode (green thin line) and the modes, denoted as L-mode (blue dashed thin line), whose real $k$'s decrease as $r$ increases. Assume that the interaction between the disks is ignored, i.e. $d=\infty$, for the moment. It is easy to see that the R-mode originates from the right disk since $R_2$ is fixed even though $r$ varies. It is clear from Fig.~\ref{fig2} (E)-(H) that the spatial distributions of R-modes exhibit strong localization only on the right disk. On the other hand, the L-mode is nothing but the WGM of the left disk since the real $k$ of the resonance should decrease as $r=R_1$ increases so does the circumference of the left disk. This is clearly shown in Fig.~\ref{fig2}(A)-(D), in which the spatial distributions are localized only on the left disk. When these two mode groups cross both R- and L-modes can coexist if no coupling between the two disks exists. In fact, the levels repel each other due to the interaction between them exhibiting the avoided resonance crossing (ARC). It is reemphasized that the modes in CNM's are strongly localized on either disk.

\begin{figure}
\begin{center}
\includegraphics[width=0.45\textwidth]{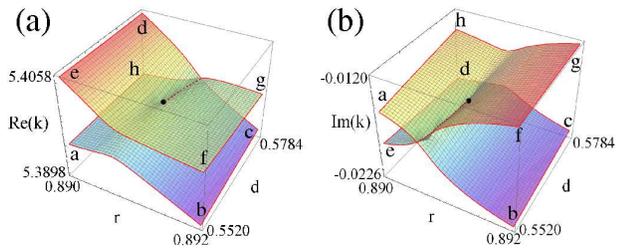}
\caption{(color online) The evolution of (a) the real and (b) the imaginary eigenvalues of the resonance modes as $r$ and $d$ are varied so that the eigenvalues are transported following $a \rightarrow b \rightarrow \cdots \rightarrow h \rightarrow a$. The EP associated with the ARC around the solid circle in the upper panel of Fig.~\ref{fig2} is located at $(r,d) \sim (0.8908,0.5572)$ denoted as a black dot.}
\label{fig3}
\end{center}
\end{figure}

\begin{figure}
\begin{center}
\includegraphics[width=0.45\textwidth]{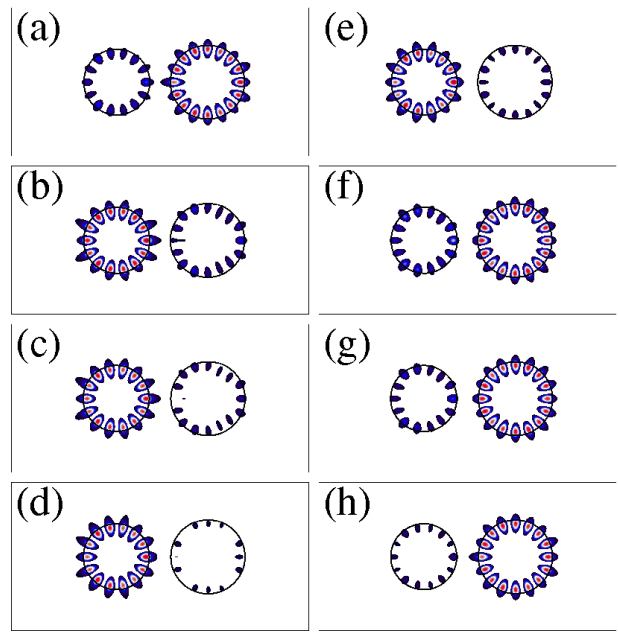}
\caption{(color online) The near field intensity patterns for a given $r$ and $d$ indicated by (a)-(h) in Fig.~\ref{fig3}.}
\label{fig4}
\end{center}
\end{figure}

\begin{figure}
\begin{center}
\includegraphics[width=0.25\textwidth]{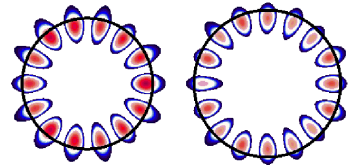}
\caption{(color online) The near field intensity pattern exactly at the EP ($k=5.3985 - i 0.01716$).}
\label{fig5}
\end{center}
\end{figure}

Let us now consider the ARC in detail. It is well known that the ARC is closely associated with the EP, a degeneracy of a non-Hermitian matrix. The ARC of the real eigenvalues usually accompany the resonance crossing (RC) of the imaginary ones, and vice versa. 
Existence of both the ARC and the RC indeed indicate the EP is located nearby. We find out the EP related to the ARC of the real $k$ around $r \sim 0.9$ indicated by a solid circle in Fig.~\ref{fig2}. One of the pronounced features of the EP is that it provides a non-trivial topology around it in a parameter space. Consider three dimensional coordinates consisting of the external parameters $r$ (x-axis) and $d$ (y-axis), and the real and the imaginary eigenvalues of the mode (z-axis) as shown in Fig.~\ref{fig3} (a) and (b), respectively. The EP takes place at $(r,d) \sim (0.8908,0.5572)$, where two complex eigenvalues coalesce. When the external parameters are continuously varied along the closed loop enclosing the EP, e.g. following the rectangular path of $(r,d)$ in Fig.~\ref{fig3}, and finally recover the initial condition, the eigenvalue does not return to the initial one. In fact, both real and imaginary eigenvalues are transported into different values, namely following $a \rightarrow b \rightarrow c \rightarrow d \rightarrow e$. In order to recover the exact original values one should encircle the EP {\em twice} following $a \rightarrow b \cdots \rightarrow h \rightarrow a$. Such behavior is also clearly shown in the spatial distribution of the corresponding modes [see Fig.~\ref{fig4} (a) - (h)].  This confirms the existence of the EP. It should be noted that at the EP the mode is equally distributed over both disks as shown in Fig.~\ref{fig5}.

\begin{figure}
\begin{center}
\includegraphics[width=0.45\textwidth]{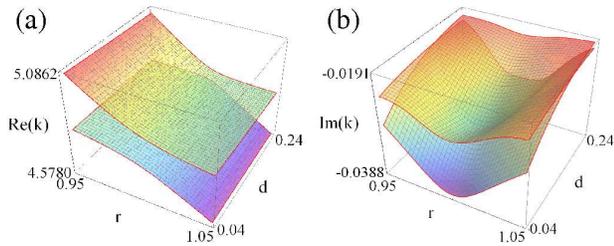}
\caption{(color online) The same as Fig.~\ref{fig3} for the ARC around the dashed circle in the upper panel of Fig.~\ref{fig2}. It shows that no EP exists.}
\label{fig6}
\end{center}
\end{figure}

One remark is in order. The ARC near $r=1$, indicated by the dashed circle in Fig.~\ref{fig2}, has nothing to do with the EP. This seems to be surprising since the ARC is known to be always associated with the EP. At $r=1$ ($R_1=R_2$) the reflection symmetry with respect to y axis induces a gap in the neighboring real eigenvalues originating from the modes with different parities, namely even and odd. Away from this point ($r=1$), the situation with a certain $r(\neq 1)$ is completely equivalent to that with $1/r$ if the left and the right disks are just interchanged. The ARC-like geometry here is thus related to the symmetry of the parameter $r$ with respect to $r=1$. Figure \ref{fig6} confirms it by showing that there is no non-trivial topology in parameter space around the ARC. Note that we search for a huge range of parameter space.

\section{Ray dynamics of coupled microdisks}

\begin{figure}
\begin{center}
\includegraphics[width=0.43\textwidth]{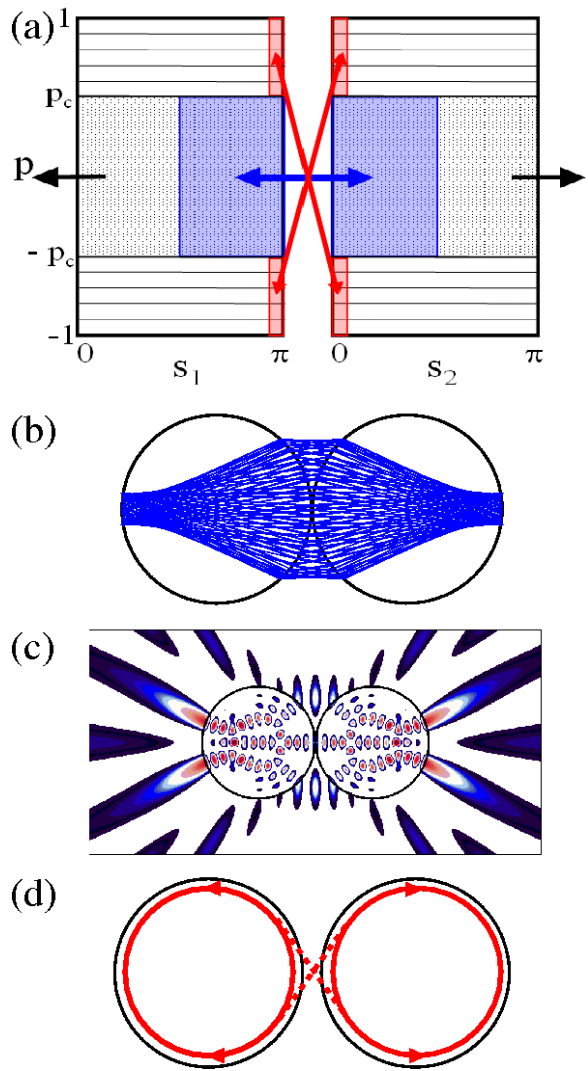}
\caption{(color online) (a) The schematic diagram of the PSOS. Without coupling ($d=\infty$) the ray dynamics is simply expressed by the horizontal straight lines. With coupling the rays below the critical angle, $|p|<p_c$ refractively escape from one disk and may enter the other disk depending on $s$, which is expressed by the horizontal arrows. If $d$ is small enough, the ray above $p_c$ can directly transmitted from one disk to the other according to the FTIR, which is expressed by the x-crossed arrows. (b) The typical orbit associated with the refractive coupling for the ray below $p_c$. (c) The near field intensity pattern of a specific resonance mode ($k=10.6116 - i 0.1334$) related to (b), which has a very low Q value. (d) The schematic picture describing the coupling through the tunneling based upon the FTIR (see the text).}
\label{fig7}
\end{center}
\end{figure}

In this section, we consider the ray dynamics of the CNM. It will become clear in the next section that the ray dynamics plays a crucial role in understanding the unidirectional emission of the CNM. The so-called Poincare surface of section (PSOS), where the angle of incidence and the position of the ray at collision on the surface are plotted in two dimensional phase space, has been used to investigate the ray dynamics of a closed cavity \cite{Rei92}. Because we have two disks, we need two PSOS's whose coordinates are provided in Fig.~\ref{fig1}. Note that due to the reflection symmetry with respect to x axis only a half of it is plotted. The ray dynamics in each disk is trivial since the angular momentum is conserved, so that the PSOS exhibits sets of horizontally straight lines. If the openness of the system is considered, the ray whose angle of incidence is smaller than the critical angle can escape from one disk and may enter the other disk. However, this is true only for the ray below the critical angle as schematically shown in Fig.~\ref{fig7} (horizontal arrows). It means the mode formed based upon this mechanism should have extremely low Q value [see Fig.~\ref{fig7} (b) and (c)]. In ray dynamics such a lossy mode formed below the critical angle is completely decoupled from the WGM-like modes formed above the critical angle.

There exists another coupling mechanism between two disks. The total internal reflection is no longer valid if the distance between the two disks is small enough, which is called as the {\em frustrated} total internal reflection (FTIR) or the optical tunneling effect \cite{Zhu86,Cha05}. It opens the possibility that the rays with the angle of incidence larger than the critical angle are inter-mixed over both disks. The inter-disk transmission probability based upon FTIR is very sensitive to $d$ so that it efficiently occurs only around the closest point of the two disks. Note that in our case $d = 0.2$ is small enough to allow considerable inter-disk transmission (the transmission probability is approximately 0.8). The clockwise (counterclockwise) rotating ray in the left disk can be transmitted into the counterclockwise (clockwise) rotating ray in the other disk, and vice versa so that this mechanism is schematically expressed by the x-crossed arrows in Fig.~\ref{fig7}(a). It is emphasized that any high Q mode should originate from the rays above the critical angle.

\section{Unidirectional emission in coupled microdisks}

\begin{figure}
\begin{center}
\includegraphics[width=0.45\textwidth]{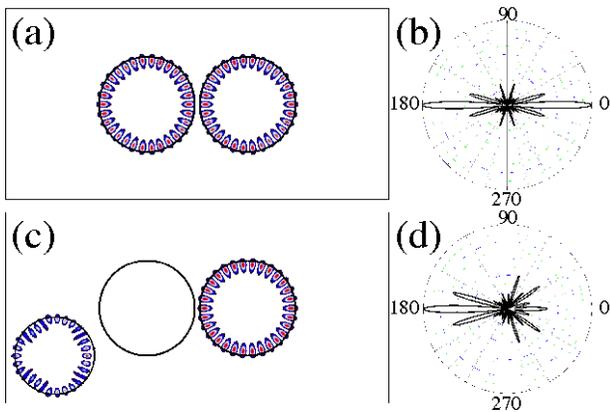}
\caption{(color online) (a) The near and (b) the far field patterns of the resonance mode ($k=9.3592 - i 0.002637$) for the coupled identical microdisks with $(r,d)=(1,0.1)$, which exhibits the bidirectional emission. (c) and (d): The same as (a) and (b), respectively, of the mode ($k=9.3398 - i 0.002588$) for the CNM with $(r,d)=(1.012,0.1)$, where the unidirectional emission takes place. The inset shows the magnified intensity pattern of the left disk, in which the distribution is localized along the diamond-like orbit.}
\label{fig8}
\end{center}
\end{figure}

It has been found that in coupled identical microdisks the bidirectional emission is obtained when the size of the disks is not only large enough but the distance between the two disks is also small enough compared with the wavelength of the mode with the appropriate parameters chosen \cite{Ryu06} as shown in Fig.~\ref{fig8} (a) and (b). Recently, the similar directional emission has been demonstrated experimentally in coupled ring lasers \cite{Sha08}. One then expects that in principle the {\em unidirectional} emission can be achieved if non-identical disks are used so that the reflection symmetry with respect to y axis is broken. It is shown in Fig.~\ref{fig8}(d) that with only slight asymmetry in radii, namely $r=1.012$, the considerable unidirectional emission takes place. In this section we explain the physical origin of the unidirectional emission in CNM by considering the ray dynamics based upon FTIR.

Figure~\ref{fig8}(c) clearly shows that the typical mode of the CNM exhibits strong localization on one disk as discussed in Sec.~II. In this case the unidirectional emission is achieved as shown in Fig.~\ref{fig8}(d). It is pointed out that even though the wavefunction is localized on the righthand side, the emission is mainly directed to the left. It leads us to investigate more carefully the wavefunction of the left disk. If one magnifies the intensity of the mode in the left disk the diamond-like pattern appears [see the inset of Fig.~\ref{fig8}(c)], which is probably ascribed to the unidirectional emission.

\begin{figure}
\begin{center}
\includegraphics[width=0.44\textwidth]{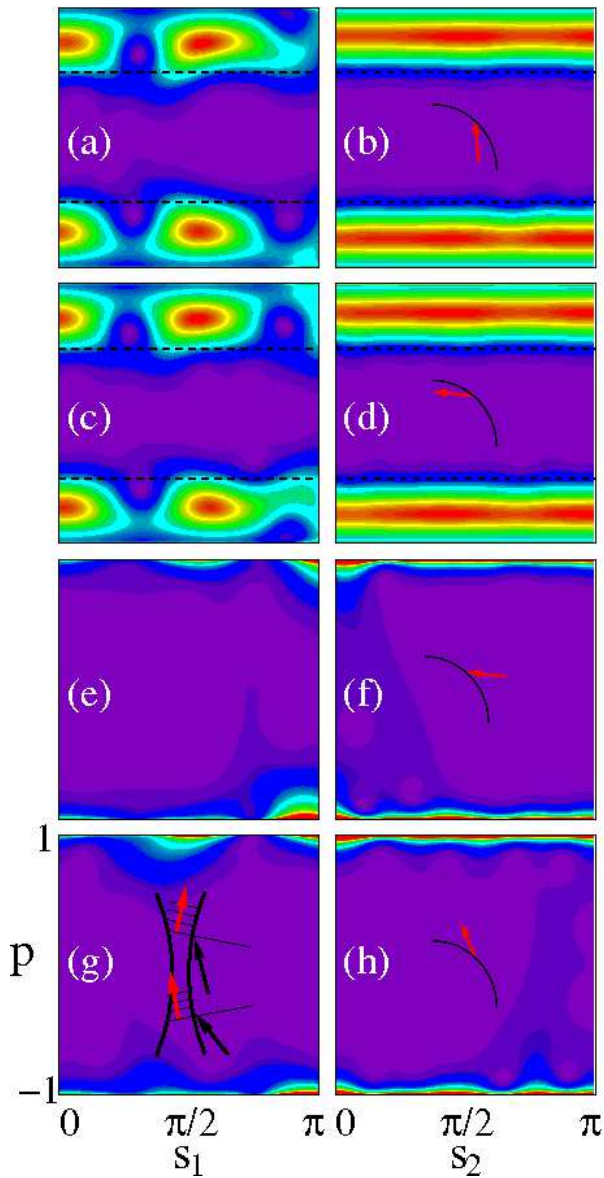}
\caption{(color online) The generalized Husimi distribution functions of the resonance mode shown in Fig.~\ref{fig8}(c); (a) $H_{(L,in)}^{inc}$, (b) $H_{(R,in)}^{inc}$, (c) $H_{(L,in)}^{em}$, (d) $H_{(R,in)}^{em}$, (e) $H_{(L,out)}^{inc}$, (f) $H_{(R,out)}^{inc}$, (g) $H_{(L,out)}^{em}$, and (h) $H_{(R,out)}^{em}$. The inset in (b), (d), (f) and (h) schematically show how each generalized Husimi function is constructed. The dashed horizontal line represents the critical angle. The inset in (g) shows the schematic picture describing the horizontal deviation between the theoretical expectation and the direct numerical result observed in Fig.~\ref{fig10}.}
\label{fig9}
\end{center}
\end{figure}

In order to get more intuition on the diamond-like pattern, we obtain the so-called Husimi distribution function, which corresponds to quantum mechanical version of the PSOS of the ray \cite{Hus40,Cre93}. Since a microcavity is an open system, the ray at collision on the boundary can tunnel out or only partially be reflected. For a given wavefunction four possible Husimi distributions can thus be constructed  \cite{Hen03}, i.e. the probability distribution of the ray either inside or outside cavity, and either incident to or emitting from the boundary at a given position of the collision. Figure~\ref{fig9} shows these four possible generalized Husimi functions, namely $H^{inc}_{X,in}$ (incident and inside), $H^{inc}_{X,out}$ (incident and outside), $H^{em}_{X,in}$ (emitting and inside), and $H^{em}_{X,out}$ (emitting and outside), where $X$ is $L$ or $R$ representing the left or the right disks, respectively.

In the right disk, the distribution looks simple [see Fig.~\ref{fig9} (b) and (d)]; it is not only uniformly distributed along the boundary $s_2$, but also located above the critical angle. In fact, this is a typical shape of a WGM. In the left disk, however, two strongly localized peaks are observed [see Fig.~\ref{fig9} (a) and (c)]. From the position of the peak $p \sim 0.73$, one can see that the angle of incidence $\chi \sim 46.9^\circ$ implying the localized peaks almost correspond to the square orbit. It is mysterious that no peak exists around $s_1=\pi$ since the square consists of four vertices. We will explain it later. Even though the probability distribution is mostly localized above the critical angle, the probability below it, however small it is, determines the direction of emission \cite{Schewefel04,SYLee05,Shinohara06,SBLee07}. Around $s_1 \sim \pi/2$ in Fig.~\ref{fig9}(a) the distribution has non-negligible overlap with the region below the critical angle, where the emission occurs. It is shown in Fig.~\ref{fig9}(g) that the emission also takes place around $s_1 \sim \pi/2$ tangentially, i.e. $p \sim 1$, leading to the unidirectional output. Therefore, the unidirectional emission originates from the tail-like structure of the central probability peak localized around $s_1 \sim \pi/2$ in Fig.~\ref{fig9}(a).

The most important question remains; why the central probability peak exists in the left disk or in other words why the square-like orbit appears in the left disk. The clue comes from the fact that there is no peak around $s_1 \sim \pi$ where the two disks are closest. When the distance between the two disks is small compared with the wavelength of the mode, the ray no longer exhibits considerable reflection from the boundary, but rather tunnels into the other disk according to the FTIR as explained above. Thus there is no reason that the intensity should be large {\em exactly at} the boundary so that no meaningful probability is identified in Fig.~\ref{fig9} (a) and (c). Recall that the rays colliding at the boundary and thus reflected from it are recoded in the PSOS. As a matter of fact, a rather higher intensity is located slightly inside the boundary at $s_1 \sim \pi$ as shown in the inset of Fig.~\ref{fig8}(c). Let us recall that the probability is mostly localized in the right disk. Without the left disk one expects uniform output emission. The FTIR induced by the left disk effectively introduces {\em a hole} on the boundary of the right disk around $s_2 \sim 0$. Through this hole the ray is transmitted mainly from the right to the left disk since the mode is mostly localized on the right disk. Figure~10 shows the probability distribution of the ray transmitting from the right, where the angle of incidence is well-defined as shown in Fig.~\ref{fig9}(b), to the left disk and colliding at the boundary of the left disk for the first time. It fits the Husimi function very well and explains the central probability peak around $s_1 \sim \pi/2$ in Fig.~\ref{fig9}(a). These rays at the next collision are mapped into the distribution centered around $s_1 \sim 0$.

\begin{figure}
\begin{center}
\includegraphics[width=0.45\textwidth]{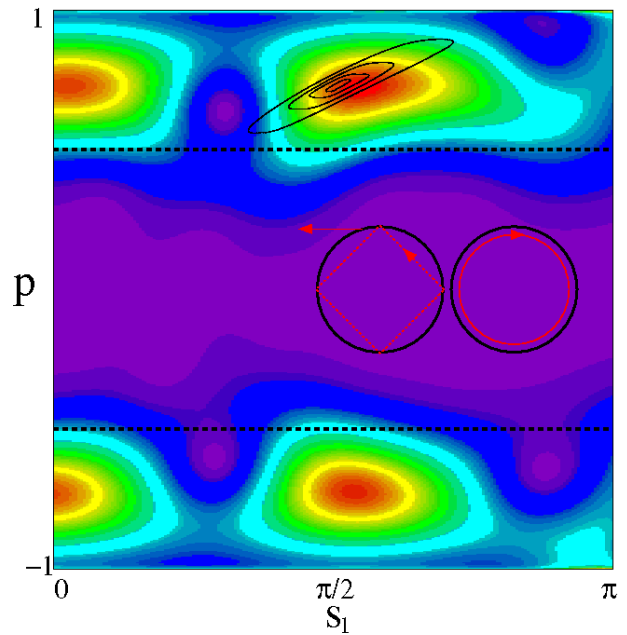}
\caption{(color online) The contours superimposed over Fig.~\ref{fig9}(a) represents the distribution of the ray expected from the theory explained in the text. The inset schematically shows how the unidirectional emission occurs in the ray picture.}
\label{fig10}
\end{center}
\end{figure}

It is noted in Fig.~\ref{fig10} that the distribution expected from the ray dynamics based upon the FTIR is slightly shifted from the Husimi function horizontally. It might be explained by considering the curved boundary instead of the flat boundary. According to the curved surface the transmission probability of the ray incident to $s_2 \sim -\delta$ may be greater than that to $s_2 \sim +\delta$ for a clockwise rotating rays in the right disk, where $\delta$ represents a certain small angle, since the former has a rather smaller incident angle inside the gap between the two disks. This situation is clearly presented in the inset of Fig.~\ref{fig9}(g). It thus induces a certain shift of the position of the peak on the boundary of the left disk. See Appendix for a possible shift for the vertical direction.

The unidirectional emission in the CNM is understood in the following way. The mode of the CNM exhibits highly asymmetric pattern; the probability distribution of the mode is localized on only one disk where the resonance condition is precisely satisfied. When the distance between the two disks is small enough the total internal reflection is no longer valid around the closest position of the two disks. It effectively generates a hole around that position. The rays are then transmitted through this hole from one disk, where the distribution is mostly localized, to the other disk. When some conditions are fulfilled, e.g. an appropriate angle of incidence of the transmitted ray in the other disk depending on the index of refraction, the unidirectional emission can take place. In this sense the unidirectional emission is not always guaranteed in a general CNM.

\section{Summary}

We have investigated two coupled microdisks with non-identical radii focusing on the parametric evolution of energy levels and the unidirectional far-field emission. It is found that the resonance modes exhibit highly asymmetric nature except the region around avoided resonance crossing; they are strongly localized on either disk depending on the external parameters. When the distance between the two disks is small enough the transmission probability near the closest position is dramatically enhanced so that the ray can be almost freely transmitted between the disks. Indeed the ray is transported from the disk, in which the mode is mostly localized, to the other, which may generate a special shape of the orbit in the other disk leading to the unidirectional emission. We believe our scenario to explain the unidirectional far-field emission may be applied to various similar class of problems.

\section*{Acknowledgment}

We would like to thank Jan Wiersig and Julia Unterhinninghofen for useful discussion. This work was supported by KRF Grant (2008-314-C00144). SYL was supported by the BK21 program.

\section*{Appendix: The angular deviation of the transmitted wave in the frustrated total internal reflection}

One might expect a vertical deviation of the peaks between the simple ray expectation and the direct numerical calculation in Fig.~\ref{fig10}. As far as the FTIR is concerned in a simple ray picture with flat boundaries, the ray tunnels through the gap and continues to travel with the angle of incidence conserved. This is not true in wave picture due to both unavoidable angular uncertainty and angular dependence of the transmission probability, leading to decrease of $p$ in the left disk \cite{Cou64}. In refractive escape, a similar phenomenon is called as the Fresnel filtering \cite{Rex02}. However, we will show that no noticeable vertical deviation has been observed here because the expected decrease of $p$ is too large compared with the observed deviation.

\begin{figure}
\begin{center}
\includegraphics[width=0.48\textwidth]{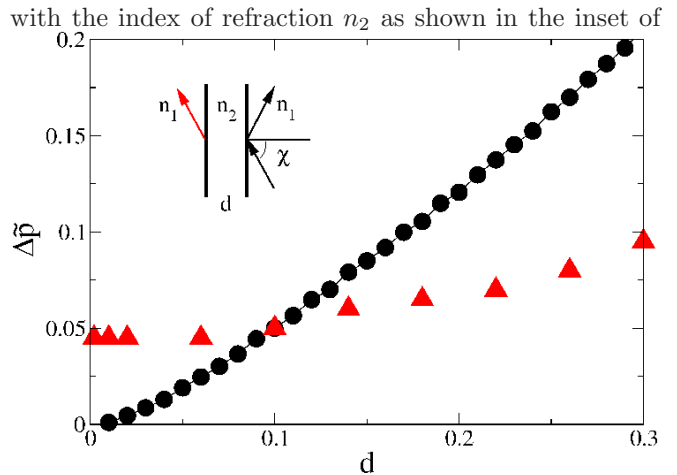}
\caption{(color online) The difference between the maxima of $p$ of the incident and the transmitted wave packets as a function of the thickness $d$ for the expectation from the planar interface (the circles) and the direct numerical calculation (the triangles).}
\label{fig11}
\end{center}
\end{figure}

Let us consider a dielectric with the index of refraction $n_1$ separated by a gap of the width $d$ filed with a dielectric with the index of refraction $n_2$ as shown in the inset of Fig.~\ref{fig11}. The transmission probability $T$ of the plane wave with the angle of incidence $\chi$ through this gap is given as \cite{Cou64}
\begin{equation}
T = \frac{1}{\alpha \mathrm{sinh}^2 y + 1}\label{a_1}.
\end{equation}
Here $y = 2 \pi d\sqrt{n^2 p^2 -1}/\lambda$ and $\alpha = (n^2 -1)^2/[4n^2 (1-p^2) (n^2 p^2 -1)]$ with $p=\sin\chi$ and $n=n_2/n_1$, where $\chi$ and $\lambda$ represent the angle of incidence and the wavelength of the incident plane wave, respectively.

As a matter of fact the incident wave in our case comes from the right disk where it has a well-defined angular distribution as shown in Fig.~\ref{fig9}(b). The angular distribution of the transmitted wave is then obtained directly from that of the incident wave multiplied by the transmission probability $T$. Since $T$ is a monotonically decreasing function of both $d$ and $p$, the angular distribution of the transmitted wave is shifted to the direction that $p$ decreases.
%Such an angular shift of the transmitted wave due to the angle-dependent transmission (or refraction in some cases) %probability is called as Fresnel filtering effect.
Figure~\ref{fig11} shows the decrease of the maximum $p$ induced by such a shift as a function of $d$ (the circles). It is mentioned that the incident angular distribution is given by the Husimi function in Fig.~\ref{fig9}(b) under the assumption of the planar interface. The triangles in Fig.~\ref{fig11} represent the decrease of the maximum $p$ obtained from the actual distribution of the mode, i.e. directly from Fig.~\ref{fig9} (a) and (b). It shows considerable discrepancy, implying the aforementioned shift is too big to explain the observed tiny deviation. Possibly other geometrical effects such as curved boundaries and a different size of the two disks might compensate the shift.

\end{document}